# Study of Proton Transfer Reaction Dynamics in Pyrrole 2-Carboxyldehyde

**Papia Chowdhury*, Rakesh Kumar, Prakash Chandra Sati and Anirban Pathak**

Department of Physics and Material Science & Engineering, Jaypee Institute of Information Technology University, Noida 201307, Uttar Pradesh, India

* Corresponding author, Fax: +91 120 2400986

**E-mail:** papia.chowdhury@jiit.ac.in

## Abstract

Photophysical and photochemical dynamics of ground state and excited state proton transfer reaction is reported for Pyrrole 2-Carboxyldehyde (PCL). Steady state absorption and emission measurements are conducted in PCL. The theoretical investigation is done by using different quantum mechanical methods (e.g. Hartree Fock, DFT, $MP_2$, CCSD etc.). The reaction pathway and two dimensional potential energy surfaces are computed in various level of theory. A transition state is also reported in gas phase and reaction filed calculation. It is established that PCL forms different emitting species in different media. A large Stokes shifted emission band, which is attributed to species undergoing excited state intramolecular proton transfer, is observed in hydrocarbon solvent. Intermolecular proton transfer is observed in hydroxylic polar solvent. Experimental observations yield all possible signatures of intramolecular and intermolecular proton transfer in excited state of PCL. The origins of these signatures have been explained successfully using corresponding quantum mechanical theories.

## Introduction

Transfer of proton from one moiety of a molecule to another moiety not only involves the exchange of charge between reactants, but it also results in transport of mass. Thus the process of proton transportation may be viewed as a realistic simple model for chemical and physical transformation [1]. The pioneering works of Weller [2-4], Weber [5] and Förster [6,7] have inspired physicists and chemists to study this phenomenon in detail with different perspectives [8-10]. For example the proton transfer process in methyl salicylate [2, 10-13] and its related compounds such as o-hydroxybenzaldehyde [14-16], o-hydroxyacetophenone [9,17], pyrrole-2-carboxylic acid [18], 2-hydroxy 1 naphtaldehyde [19,20] etc. have been studied in detail with appropriate attention towards potential applications in energy/data storage devices, optical switching [14,21], Raman filters and hard scintillation counters [22], polymer photostabilizers [23], conducting wires [24] etc. From these studies it is strongly established that a bifunctional organic molecule having acidic and basic groups in close proximity may undergo an excited state intramolecular proton transfer (ESIPT) [25]. This is expected, as the charge distribution in excited state is different from that in ground state. Consequently the dynamic and equilibrium properties of the excited state differ considerably from those of the ground state. This provides an easy pathway for the transfer of charge (electron or proton) from one moiety to other. That is why charge transfer process easily takes place in excited state. This aspect of proton transfer process is discussed in several earlier works [3, 4, 26-28]. Environment also plays a crucial role in this charge transfer phenomenon.

Generally most of the studies on intramolecular proton transfer are focused on non-aqueous solutions [19]. But the photo physics of ESIPT process in aqueous solution is of more industrial interest than that in non-aqueous solvents [19]. However these studies are very complicated as the interaction of the molecule with water and other hydroxylic solvents lead to many processes that do not occur in non-aqueous solutions [20]. This fact has motivated us to study the effect of hydroxylic solvent on ESIPT of PCL. The ESIPT processes take place through an intramolecular hydrogen bond, which is usually formed in a six or five member ring. Since PCL is a heterocyclic molecule having five member ring with two moieties, it is capable of establishing an intramolecular hydrogen bond. Consequently there exists a probability of observing photo-induced intramolecular proton transfer reaction in PCL. This fact along with the potential applications of PCL in tobacco industry[28a-d] and that of proton transfer process in design of molecular wire, data storage[28e,f], fuel cell and in evolution processes in DNA[28g-i] have also motivated us to study the possibility of observing proton transfer in PCL which is a derivative of DNA base. In our previous work [29] we have shown the existence of an intramolecular hydrogen bonding between $O_{11}$…..$H_{12}$ of formyl hydrogen and pyrrole nitrogen of PCL (scheme: 1). This had strongly indicated the possibility of proton transfer. This possibility had been supported by other observations like the calculated difference in enthalpy ($\Delta H$= 2.32 Kcal/mol) and that in free energy ($\Delta G$= 1.90 Kcal/mol) for the molecule. These results indicated that the normal form of the molecule in the ground state is moderately more stable than that in the excited state. Further, the optimized structure of PCL shows angle between $C_1$-$C_{10}$-$O_{11}$ changes from 123.35° (in ground state) to 119.83° (in excited state). The distance between acidic and basic part of PCL (i.e. $H_{12}$-$O_{11}$) is theoretically observed to decrease from (2.60 Å) in ground state to (2.31 Å) in the excited state. It is also observed that the inter ring bond angle between $H_{12}$-$N_5$-$C_1$ decreases from 122.3 ° (in ground state) to 118.8° (in excited state). All these changes in bond length, inter ring bond angle etc. are consistent with the mechanism of proton transfer. This prompted us to predict the existence of intramolecular or intermolecular charge transfer in the excited state between $H_{12}$………$O_{11}$. The prediction was supported by the large change in dipole moment (3.10 Debye in ground state and 3.49 Debye in excited state). Further it was observed that the change in dipole moment is more in water medium and consequently possibility of intermolecular proton transfer is more in water medium. All these theoretical and experimental (UV-VIS absorption, FTIR absorption) results obtained in our previous work [29] had only indicated the existence of intramolecular hydrogen bond and chance of intramolecular proton transfer in the excited state. But the results were not confirmative. The present work aims to confirm the earlier prediction of ours. Further, the charge transfer reaction dynamics in PCL were not studied in the earlier works and consequently no generally accepted molecular mechanism has been emerged till now. This has motivated us to theoretically establish the reaction pathways for the normal proton transfer reaction. It is attempted by characterizing the potential energy surface (PES), the geometry and energy of the stationary points (reactants (R), transition structure (TS) and product (P)). The effects of different kind of solvents on the reaction dynamics of proton transfer process in PCL is also studied in the present work and the emission spectra of PCL have been obtained in different media (e.g. Methyl cyclohexane, Methanol, water etc.). In this context we would like to note that the experimental investigations incurred in the present work have yield all the familiar signatures of proton transfer. The origins of these signatures have been explained successfully using corresponding quantum mechanical theories.

In this paper we have studied both theoretically and experimentally the excited state characteristics of PCL in different environment. The effect of environment on the dynamics proton motion and the interaromatic ring rotation are studied. In the next section we have briefly described the theoretical and experimental methods followed in the present work. In section 3 we have discussed the experimental results of the present work and in section 4 we have provided quantum mechanical explanations of the experimental findings. Finally section 5 is dedicated to conclusions.

**Materials and Theory**

**Method1.** The gas phase ground state geometry of PCL is fully optimized with and without symmetry constraints using TDDFT with the Becke3LYP hybrid functional, $MP_2$, CCSD and HF theory with the 6-311G, 6-311G (d, p), 6-311G (2d,2p) basis sets respectively. All the basis sets are used as implemented in Gaussian 03 [30,31]. Configuration interaction with singlets (cis) has been computed in order to optimize the first and second excited singlet state ($S_1$, $S_2$) [32]. For DFT calculations the optimizations are performed with symmetry constraints. Long range effects of the solvent medium are taken into account by means of a dielectric continuum represented by the Onsager model [33, 34]. The default options for the self–consistent field (SCF) convergence and threshold limits in the optimization are used. DFT calculations are performed on the gas phase optimized geometry of the ground state ($S_0$) [35]. For CCSD methods we have included doubles and optional triples terms. Optimized geometries of PCL are used to calculate gas phase vertical excitation energies in CIS method with 6-311G basis set.

Presence of solvent introduces correlation energy which may affect the proton transfer reaction dynamics. The effect is studied in detail and the results are presented together with the structure and energy profile along with the reaction path in the gas phase and in solution. The PESs have been calculated in detail to ensure that all relevant stationary points have been located and properly characterized. The exact location of the TS is obtained by using an algorithm [36-38] in which the coordinates describing the system are separated into two sets.

**Materials2.** PCL, PCA, Methyl cyclohexane (MCH), Methanol ($CH_3OH$) and Potassium Bromide (KBr) are purchased from Aldrich. MCH, $CH_3OH$, KBr are used as supplied but only after checking the purity fluorimetrically in the wavelength range of interest. Deionized water (Millipore) is used to measure the absorption and emission spectra. The UV-VIS absorption spectra at 300K are recorded with a Perkin Elmer spectrophotometer (Model Lamda-35) with varying slit width. The UV-VIS emission spectra at 300K are recorded with Perkin Elmer spectrophotometer (Model Fluorescence-55) with varying slit width. The concentration of PCL used in solution phase experiments is of the order of $10^{-5}$ M.

**Results and Discussion**

Absorption bands of PCL are observed at 288 nm and at 245 nm in hydrocarbon solvent (figure-1). These bands have been little red shifted in the polar and hydroxylic solvents (as shown in figure-1). With the help of theoretical calculation and FTIR data the existence of intramolecular hydrogen bonding between $O_{11}$…..$H_{12}$ of formyl hydrogen and pyrrole nitrogen of same molecule is established. We have already predicted in [29] that there is a strong chance of intramolecular proton transfer between these two groups, provided the molecule is in the excited state. In the present work we have obtained the emission spectra of the molecule in different media (e.g. methyl cyclohexane, methanol, water etc.). We have tested both the hydrocarbon and hydroxyl medium. In the hydrocarbon medium the molecule is free to move as hydrocarbon medium is nonpolar in nature so it will not interact with molecule directly. In hydroxyl medium there may be influence of external hydrogen bonding [39]. By exciting the molecule at 288 nm in MCH we have obtained a single strong wide band at ~366 nm (figure-2). By exciting the same sample at 245 nm the band is observed at the same position but with lower intensity. But in methanol, exciting the molecule at 288 nm we have observed dual fluorescence with a sharp band at ~ 322 nm and a weak band at ~380 nm (figure-3). Exciting the PCL at 245 nm, a dual fluorescence is observed with a strong band at ~ 380 nm and a weak band at ~322 nm. Since MCH is a nonpolar medium so there is no chance of any environmental effect. Most of the PCL molecules in MCH may remain in the cis (E) conformeric form since the interaction between $N_5$-$H_{12}$ and $C_{10}$-$O_{11}$ in this medium is strong [29]. This leads to a high probability of intramolecular proton transfer in MCH solvent. In MCH this large Stokes shifted (~8000 $cm^{-1}$) band at 366 nm may be attributed to zwitterionic form, which is created due to intramolecular proton transfer between the acidic and basic moiety of the same molecule. While in methanol, the high polarity of the medium and existence of hydroxyl group may cause breaking of some intramolecular hydrogen bonding and creation of some intermolecular hydrogen bonding. This may lead to the rotation of basic -CHO group to give the trans configuration of the molecule. So in polar medium there is a high possibility of existence of normal trans (b) [29] conformer. It should give a peak at lower wavelength side of spectrum. The same is observed here. It is interesting to note that in pure hydroxylic solvent water a new behavior of the molecule is obtained (figure-4). To be precise, if PCL is excited at 280 nm then in the emission spectra a new band appears at ~ 450 nm along with a band at ~372 nm. This new band is totally absent in hydrocarbon medium and other polar medium like methanol. So the new band at 450 nm may be assigned due to intermolecular proton transferred form of the molecule.

Variation of pH of the medium considerably affects the spectral properties of PCL. Gradual addition of base (sodium ethoxide) in water solution of PCL results in red-shift of the above mentioned new band. The band, whose intensity is found to increase with higher pH of base, is now observed at ~ 450 nm (for pH=12, see figure-4). Excitation spectra corresponding to 450 nm emission band in basic medium correlates well with the absorption spectrum of PCL in water and in presence of base. Thus the emission band at 450 nm and 372 nm are assigned to the anion due to intermolecular proton transferred form and zwitterion due to intramolecular proton transferred form respectively. Anion formation due to subtraction of proton by addition of base as shown in scheme: 2. Here the large Stokes shifted fluorescence of the anion is attributed to its capability of resonance stabilization by extensive delocalization of the negative charge of the π-electrons of the nitrogen present in pyrrole. Base has no effect on the emission spectra of PCL in non-polar solvents. This is probably due to the presence of strong intramolecular hydrogen bonding in PCL in non-polar solvents, which can not be ruptured even by strong base. This observation confirms the existence of anionic form of the molecule which occurs due to intermolecular proton transfer from molecule to solvent. At this point we would like to note that after the important work of Weller [4] on proton transfer dynamics many people have started work in this topic sometime with the name proton transfer and sometime with the name H-atom transfer. But these two are not synonyms [1]. In case of H-transfer the hydrogen atom really transfers from one molecule to another with a redistribution of electronic charge (scheme:3). In proton transfer process the product is zwitterionic tautomer where charge transfer occurs within the molecule but overall charge of the molecule remains same.

The opposite effect is observed in presence of acid in polar medium. With increasing acid concentration in water medium the peak intensity of 372 nm increases abruptly (figure-5). This effect also supports the existence of anionic form of the molecule at 450 nm. Addition of dilute $H_2SO_4$ in water solution of PCL results in increase in intensity of lower wavelength ESIPT band (~372 nm) with a decrease in higher wavelength band (~450 nm). Addition of acid results in the rupture of intermolecular hydrogen bonding present between solute and solvent in hydroxylic and polar protic solvents as shown in scheme: 2. Increase in acid concentration results in complete disappearance of the higher wavelength band. Addition of acid in polar solution of PCL also results in increase in intensity of band at ~372 nm. So addition of acid favors the zwitterionic formation. It is possible that at high acid concentration cation formation occurs which leads to appearance of emission band around 372 nm.

## Quantum Chemical Calculations

The conformational structures of PCL molecule are obtained in the electronic ground state by HF method and DFT by using B3LYP hybrid functional with 6-311G(d,p) basis set. The calculated structures are named as follows: the reactant (normal PCL) is designated as E, product (zwitterionic structure of PCL) is designed as Z*, the transition structure is labeled as TS and the anionic structure of PCL is designated as A (scheme: 4). The structural parameters for the cis (E) and trans (K) form are presented in table: 1. PCL may exist in different forms as the aldehyde group attached to five membered ring may be rotated to different orientation. All these possible structures are optimized using HF, DFT, CCSD, $MP_2$ etc. and are shown in scheme: 1. It is observed from the calculated energy, dipole moment, enthalpy etc. that the intramolecularly hydrogen bonded closed conformeric form (E) is the most stable among these possible structures of PCL. A potential energy curve is presented in figure-6, which shows the variation of the ground state energy with $<C_1$-$C_{10}$-$O_{11}$ bond angle for cis conformer. We have not shown the potential energy curve for trans conformer because the characteristics of the potential energy curves for the trans conformer are same as that of cis conformer. For cis conformer the potential energy of the system approaches a lower value compared to that of the trans conformer. Energy difference between these two forms (cis and trans) comes out to be 82.83 Kcal/mol.

The negative charge distribution is computed by using the Mulliken charge scheme at HF and DFT level and it shows an increase in negative charge density on O atom of C=O group (-0.44 in ground state (GS) to -0.60 in excited state (ES)). Similarly negative charge density decreases on N atom of –NH group (-0.64 in GS to -0.45 in ES). This indicates the possibility of proton transfer (from $–N_5H_{12}$ group to –CHO group of PCL) which is further supported by the computed values of Gibbs free energy, enthalpy, bond length, bond angle, dihedral angle etc(table:2) [29]. The investigation on the variation of the potential energy for the proton transfer reaction requires to define a suitable reaction coordinate (RC) along which the change in energy is to be observed. In this case the obvious choice of RC is the coordinate along which the proton is transferred from $N_5$ to $O_{11}$ that is $R_{OH}$ or $R_{NH}$ distance. When we do rigid calculation (i.e. when we freeze all other degrees of freedom except the RC) the potential energy curves (PEC) for the ground and excited state show a single minimum due to E conformer (not shown in figure). But it is observed that when changes occur in RC due to the change in $N_5$-$H_6$ distance and $O_{11}$-$H_{12}$ distance, the other parameters (mainly $<C_1-N_5-H_{12}$ or $<C_1-C_{10}-O_{11}$) also undergo small changes during the process of proton transfer. Consequently the interaction energy decreases between $O_{11}$-$H_{12}$. Thus we can conclude that the rigid calculation yields an incomplete picture of the proton transfer process and the relaxed calculation may provide a clearer description of the proton transfer reaction dynamics.

The energy (in Kcal/mol) vs. variation of $N_5$-$H_{12}$ bond length ($R_{NH}$) is now plotted again with optimizing all other coordinates (relaxed calculation) (figure-7) (table:3). It is observed that the energy of the $S_0$ (obtained using DFT) and $S_1$ (obtained using TDDFT) states increases with the increase of $N_5$-$H_{12}$ bond length. Singlet ($S_0$) state shows single sharp minimum at the structure of E form in the PEC. Also the calculated $S_1$ state shows a sharp minimum energy at E* form along the reaction coordinate. But in addition to this minimum another minimum occurs for the $S_1$ state in the higher value of $N_5$-$H_{12}$ bond length. This second minimum may be attributed to the Z* form of the molecule which occurs due to intramolecular proton transfer between acidic and basic moieties of the molecule. The sharp minimum for Z* in PEC scan shows that in the excited state Z* form is stable and probably barrier less proton transfer process yields this state of the molecule. This provides a strong signature of a proton transfer reaction in the $S_1$ state of PCL. We have also calculated the oscillator strength with variation of $N_5$-$H_{12}$ bond length in the excited state during the transformation from E* form to Z* form. It is found that the calculated oscillator strength increases from E* form to the Z* form in the $S_1$ state (figure-8). Generally strength of electronic transition is expressed in terms of oscillator strength. Its ideal value is unity which is obtained when all the molecules are transferred to the higher energy state from its ground state. Higher value of *f* corresponds to larger formation of Z*. Larger formation of Z* gives larger emission intensity which corresponds to high quantum yield. Therefore, it is confirmed that on excitation PCL attains a delocalized excited state (E*), which then relaxes to the proton transferred configuration with the transfer of a proton from –NH group to –CHO group and forms the zwitterions (Z*). An DFT calculation yields energies (in Kcal/mol) for different states with variation of bond length and bond angle at aldehyde site and at pyrrole site (figure-9). The structures associated with the stationary points characterized on PES are given in the scheme: 3. The PES has been studied in detail to ensure that all relevant stationary points can be located and properly characterized. The corresponding energy profile for the process is sketched in figure-9. The selected geometrical parameters and relative energies for the stationary points along with the reaction pathways are presented in table: 3. All optimized geometries are in the planar configuration. The analysis of the energy profiles obtained at different levels of theory show that the TS controls the charge transfer process from the kinetic energy point of view. The relative change in energy of the stationary points from E to TS is obtained as 33.88 kcal/mol and that of TS to Z* is 18.82 kcal/mol. The $N_5$– $H_{12}$ and $C_{10}$-$O_{11}$ distance, corresponding to proton transfer process between $N_5$ and $O_{11}$ is the main component of the TS. The $N_5$-$H_{12}$ distance changes drastically from 0.99 Å in the reactant (E) to 1.10 Å in TS, indicating that at TS, the first protonn to be transferred from acidic hydrogen ($H_{12}$) of the pyrrole group to >C=O group of basic aldehyde moiety is at the midway point. The hydrogen bond distance of $H_{12}$-$O_{11}$ decreases from 2.05 Å to 0.95 Å in the process of transformation from E to product form (Z). The result correlates with the calculated PCL energies for the normal (E) form and the zwitterions (Z*) form at B3LYP level using 6-311G(d,p) basis set. Here all computational methods describe the reaction as an endothermic process in the range of 33.88 Kcal/mol to 18.82 Kcal/mol and the reaction pathway presents an inverted energy profile. The reaction path posses a flat region, which corresponds to the first proton transfer, is shown in figure-9. After proton transfer the reaction path takes a sharp turn and that lead to the TS. We have obtained another transition structure profile in presence of hydroxylic medium like water (scheme: 4 and figure-10). In presence of water PCL first makes an intermolecular hydrogen bond between $H_{12}$ and $O_{13}$ of water molecule. Transition structure controls the hydrogen transfer from PCL to the water molecule and as a product form we obtain the anion of PCL. Here the energy difference between initial structure and the transition structure is 15 Kcal/mol.

The balanced measure of the extent of bond formation or bond breaking along the reaction pathway is provided by the concept of bond order (BO). This theoretical tool has been used to extensively study the molecular mechanism of chemical reaction [40, 41]. The Wiberg bond indices [42] have been computed by using the natural bond orbital [43] analysis, which gives the nature of decomposition process. The progress of the chemical process at transition structure was evaluated through the following expression:

$$\%\,evolution = \frac{BO(TS) - BO(E)}{BO(Z^*) - BO(E)} \times 100 \qquad (1)$$

where E and Z* represent possible reactants/intermediates and intermediates/products, linked along a particular reaction pathway: E $\rightarrow$ TS $\rightarrow$ Z*. Scheme: 3 shows the different state of the molecular system during the transition from E to Z*. Calculated percentage of evolution at TS using the HF/6-31g(d,p) method show the % evolution of $N_5$-$H_{12}$ breaking and that of $O_{11}$-$H_{12}$ forming process are 33.34 % and 75.30 % respectively. So the intramolecular charge (hydrogen) transfer via TS is associated with an intramolecular hydrogen transfer. In this process the probability of formation of $O_{11}$-$H_{12}$ is more compared to that of $N_5$-$H_{12}$ breaking process in gas medium. Whereas in hydroxylic medium the $N_5$-$H_{12}$ breaking process is more probable (76 %) compared to the $H_{12}$-$O_{13}$ forming process (22 %). This is quite natural in a hydroxylic medium as the intermolecular proton transfer from solute to solvent is more probable.

Similarly in presence of specific reaction field like water we have calculated the PES of the molecule. Results show the existence of intermolecular hydrogen bonding between the solute and solvent. Here the suitable reaction coordinate is the bond distance between $O_{11}$-$H_{12}$. It is observed that two minima occur in two sides of the transition structure (figure-10). According to theory if stable reactant form (Z) transfers to another stable product form (A) then there must exist a transition structure (TS) between these two stable structures. Energetically TS should remain at higher energy level than that of reactant and product energy levels. So TS should have higher energy value with two minima on both side of it. Experimentally we got the existence of two bands Z* and A* in water. We predicted the reason for these bands are due to zwitterionic form (Z*) and anionic form (A*).

**Conclusions**

Since PCL is a heterocyclic bifunctional organic molecule with two moieties present in proximity, it is easy to identify PCL as a molecule which may show proton transfer reaction. Motivated by this fact we had studied the possibility of proton transfer in PCL in an earlier study [29] which had provided us some elementary signatures of proton transfer in excited state of PCL through change in dipole moment, change in bond length, change in Gibb's energy, change in bond angles etc. But none of these signatures were confirmative. Present study has confirmed the proton transfer reaction process with the help of steady state emission spectra and other observations. Further the effect of environment on the proton transfer process is studied. The study has established that PCL forms different emitting species in different media. A large Stokes shifted emission band, which is attributed to species undergoing excited state intramolecular proton transfer, is observed in hydrocarbon solvent. On the other hand, intermolecular proton transfer is observed in hydroxylic polar solvent. A transition state is also reported in gas phase and reaction filed calculation. Experimental observations (UV-VIS absorption spectroscopy, fluorescence spectroscopy) and the computational analysis conducted in the present study and FTIR data in our previous study [29] have provided us all the familiar signatures of intramolecular and intermolecular proton transfer in excited state of PCL. The origins of these signatures of intramolecular and intermolecular proton transfer have been explained successfully using corresponding quantum mechanical theories (e.g. HF, TDDFT, $MP_2$, CCSD etc.). Altogether we have provided a complete description of proton transfer reaction dynamics in PCL. This detail understanding of reaction dynamics of proton transfer in PCL is expected to provide deeper insight into the evolution process, since PCL is a derivative of DNA base. Consequently present study also opens up the possibility of similar studies in other DNA derivatives for a complete understanding of the evolution process.

**References:**

1. Douhal A, Lahmani F and Zewail A H, Proton-transfer reaction dynamics. *Chem. Phys*. (1996) **207** 477-498.
2. Weller A, Quantitative untersuchungen der fluoreszenzumwandlung bei naphtholen,Z. *Eektrochem* , (1952) **56** 662.
3. Weller A, Allgemeine basenkatalyse bei der elektrolytischen dissoziation angeregter, *Electrochem***,** (1957) **61** 956.
4. Weller A, Outer and inner mechanism of reactions of excited molecules *Discuss Faraday Soc,* (1959) **27** 28-33.
5. Weber K, Close connection between extinction of fluorescence and retardation, Z *Phys. Chem.*, (1931) 15 18.
6. Förster T,Fluoreszenzspektrum und Wasserstoffionen-konzentration *Naturwissenschaften*, (1949) **36,** 6 186-187.
7. Förster T, Electrolytic dissociation of excited molecules *Electrochem, Elektrolytische Dissoziation angeregter Molekule*, (1950) **54** 42 and 531.
8. Kasha M, Proton transfer spectroscopy. Perturbation of the tautomerization potential, *J. Chem. Soc. Faraday Trans. II,* (1986) **82** 2379-2392.
9. Peteanu L A, Mathies R A, (Resonance Raman intensity analysis of the excited-state proton transfer in 2-hydroxyacetophenone *J. Phys. Chem*., (1992) **96,** 17 6910-6916.
10. Herek J L, Pedersen S, Banares L, Zewail A H, Femtosecond real-time probing of reactions. IX. Hydrogen-atom transfer, *J. Chem. Phys ,*(1992) **97** 9046-9061.
11. Law K Y, Shoham J, Photoinduced Proton Transfers in 3,5-Di-tert-butylsalicylic Acid, *J. Phys. Chem*., (1995) **99** 12103-12108.
12. Goodman J, Brus L E, Proton transfer and tautomerism in an excited state of methyl salicylate. Proton transfer and tautomerism in an excited state of methyl salicylate,

*J Am. Chem. Soc*. ,(1978) **100** 7472-7474.

13. Aucna A U , Toribio F, Amit-Guery F, Chatalan J, Excited state proton transfer: A new feature in the fluorescence of methyl 5-chlorosalicylate and methyl 5-methoxysalicylate, *J Photo-chem*., (1985) **30** 339-352.
14. Morgan M A, Orton E, Pimentel G C, Characterization of ground and electronically excited states of o-hydroxybenzaldehyde and its non-hydrogen-bonded photorotamer in 12 K rare gas matrixes, *J. Phys. Chem*., (1990) **94,** 20 7927-7935.
15. Nagaoka S, Nagashima U, Ohta N, Fujita M, Takemura T, Electronic-state dependence of intramolecular proton transfer of o-hydroxybenzaldehyde Electronic-state dependence of intramolecular proton transfer of o-hydroxybenzaldehyde*, J. Phys. Chem,*(1988) **92,** 1 166-171.
16. Nagaoka S, Hirota N, Sumitani M, Yoshihara K, Lipezynska-Kochany E, Lwamura H, Investigation of the dynamic processes of the excited states of o-hydroxybenzaldehyde and its derivatives. 2. Effects of structural change and solvent , *J.Am. Chem. Soc*., (1984) **106,** 23 6913-6916.
17. Nishiya T, Yamauchi S, Hirota N, Baba M, Hanaxaki I, Fluorescence studies of intramolecularly hydrogen-bonded o-hydroxyacetophenone, salicylamide, and related molecules, *J. Phys. Chem.,* (1986) **90,** 22 5730-5735.
18. Sahoo D, Adhikary T, Chowdhary P, Chakravorti S, Theoretical study of excited state proton transfer in pyrrole-2-carboxilic acid, *Molecular Phys.,* (2008) **106** 1441-1449.
19. Chowdhary P, Panja S, Chakravorti S, Excited state prototropic activities in 2-hydroxy 1-napthaldehyde*, J. Phys. Chem. A*, (2003) **107** 83.
20. Chowdhary P, Chakravorti S, Efffects of micelles on excited state intramolecular proton transfer activities of 2- hydroxy 1-naphthaldehyde. *Chem. Phys Lett.,* (2004) **395** 103.

21. Kuldova K, Corval A, Trommsdorff H P, Lehn J M, Photoinduced Generation of Long-Lived Proton Transfer States: Photoinduced Proton Transfer from 2-(2‘,4‘-Dinitrobenzyl)pyridine to a Proton Cage, the [2.1.1] Cryptand, *J. Phys. Chem.* ,(1997) **101** 6850-6854.

22. Martinez M L, Cooper W C, Chou P T, A novel excited-state intramolecular proton transfer molecule, 10-hydroxybenzo[h]quinoline, *Chem. Phys. Lett*., (1992) **193** 151-154.

23. Heller H J, Blattmann H R, Some aspects of stabilization of polymers against light *Pure, Appl. Chem*., (1973) **36** 141-162.

24. Alkorta I, Elguero J, Theoretical models of directional proton molecular transport , *J. Org. Biomol. Chem*., (2006) **4** 3096-3101.

25. Formosinho S J, Arnaut L G, Excited-State Proton Transfer Reactions I. Fundamentals and intermolecular reactions, Excited-State Proton Transfer Reactions II. Intramolecular Reactions, *J. Photochem. Photobiol. A Chem.*, (1993) **75** 1-20, 21-48.

26. Itoh M, Kurokawa H, Excitation energy dependence on the intramolecular excited- state proton transfer of 3-hydroxyflavone in the vapor phase, *Chem. Phys. Lett,* (1982). **91** 487-490.

27. Ernasting N P, Mordzinsky A, Dick B, Excited-state intramolecular proton transfer in jet-cooled 2, 5-bis (2-benzothiazolyl) hydroquinone *J. Phys. Chem.* ,(1987) **91** 1404-1407.

28. Kelley R N, Schulman S G, Molecular luminescence Spectroscopy: Methods and applications part II, (Wiley, New York), (1988) 467.

28a. Fernndez D S B, Esteruelas E, Munoz A M, Estrella C and Sanz M ,Volatile Compounds in Acacia, Chestnut, Cherry, Ash, and Oak Woods, with a View to Their Use in Cooperage, *J. Agric. Food Chem.,* (2009) ***57*** (8)**,** pp 3217–3227 DOI: 10.1021/jf803463.

28b. Natali N, Fabio C and Claudio R ,Characterization of Volatiles in Extracts from Oak Chips Obtained by Accelerated Solvent Extraction (ASE), *J. Agric. Food Chem.*, (2006), ***54*** (21), pp 8190–8198DOI: 10.1021/jf0614387.

28c. John C L and Alford E D, Volatile constituents of perique tobacco, *Electronic journal of Environmentsl agricultural and food chemistry*, (2005) **4**, 2.

28d. Lu X, Cai J, Kong H, Wu M, Hua R , Zhao M , Liu J, and Xu G**,** Analysis of Cigarette Smoke Condensates by Comprehensive Two-Dimensional Gas Chromatography/Time-of-Flight Mass Spectrometry I Acidic Fraction, *Anal. Chem.* (2003*), **75*** (17), pp 4441–4451DOI: 10.1021/ac0264224.

28e. Yen Y S, Hsu Y C, Lin J T, Chang C W, Hsu C P and Yin D J, Pyrrole-Based Organic Dyes for Dye-Sensitized Solar Cells, *J. Phys. Chem. C*, (2008) **112** (32), pp 12557–12567 DOI: 10.1021/jp801036s.

28f. Jolicoeur, Benoit, Lubell, William D **,** Prodigiosin synthesis with electron rich 2,2′-bipyrroles, *Canadian Journal of Chemistry,* (2008) Volume **86**, Number 3, 1 March , pp. 213-218(6).

28g. Bergamaschi G, Bonardi A, Leporati E, Mazza P, Pelagatti P, Pelizzi C , Pelizzi G, Rodriguez M C , Zani A F**,** Organotin complexes with pyrrole-2-carboxaldehyde monoacylhydrazones. Synthesis, spectroscopic properties, antimicrobial activity, and genotoxicity, *Journal of Inorganic Biochemistry,* (1997) Volume **68**, Issue 4, December, Pages 295-305.

28h. Bacchi, Bonardi A, Carcelli M, Mazza P, Pelagatti P, Pelizzi C, Pelizzi G, Solinas C and Zani F, Organotin complexes with pyrrole-2,5-dicarboxaldehydebis(acylhydrazones). Synthesis, structure, antimicrobial activity and genotoxicity. *Journal of Inorganic Biochemistry,* (1998)*,* Volume **69**, Issues 1-2, 1 February, Pages 101-112 .

28i. Njoroge F G, Fernandes A A , Monnie V M ,3-(D-*Erythro*-Trihydroxypropyl)-1-Neopentyl Pyrrole-2-Carboxaldehyde. A Novel Nonenzymatic Browning Product of Glucose , *Journal of Carbohydrate Chemistry*, (1987) Volume **6** , Issue 4 December, pages 553 – 568.

29. Kumar N, Chakravorti S , Chowdhury P, Experimental Investigation by UV-VIS and IR Spectroscopy to Reveal Electronic and Vibrational Properties of Pyrrole-2-Carboxyldehyde: a Theoretical Approach*, Journal of Molecular structure,* (2008) **891** 351-356.

30. Beche A D, Density-functional thermochemistry. III. The role of exact exchange J. *Chem .Phys*., (1993) **98** 5648-5652.

31. Pople J A , Gaussian 98, Gaussian Inc, Pittsburgh, PA, (1998).

32. Foresman J B, Head-Gordon M, Pople J A, Frich M J, Toward a systematic molecular orbital theory for excited states, *Phys. Chem*., (1992) **96** 135-149**.**

33. Alina D T, Slawomir G J, Dorota R B, Tomasz M, Jerzy L, Pyrrole-2-carboxylic Acid and Its Dimers: Molecular Structures and Vibrational Spectrum, *J. Phys. Chem. A*, (2002) **106** 10613-10621.

34. Onsagar L, Electric moments of molecules in liquids, *J. Am. Chem. Soc.*, (1936) **58** 1486-1493.

35. Stratmann R E, Scuseria G E, Frich M J*,* An efficient implementation of time-dependent density-functional theory for the calculation of excitation energies of large molecules, *J. Chem. Phys. Lett*., (1998) **109** 8218.

36. Tapia O, Andress J, A simple protocol to help calculate saddle points. Transition-state structures for the Meyer—Schuster reaction in non-aqueous media: An ab initio MO study, Chem. *Phys. Lett*., (1984) **109** 471-477.

37. Andress J, Moliner V, Safont V S, Theoretical kinetic isotope effects for the hydride-transfer step in lactate dehydrogenase, *J. Chem. Soc., Faraday Trans*., (1994) **90** 1703-1707.

38. Tapia O, Andress J, Safont V S, Enzyme catalysis and transition structures *in vacuo*. Transition structures for the enolization, carboxylation and oxygenation reactions in ribulose-1,5-bisphosphate carboxylase/oxygenase enzyme (Rubisco), J. Chem. Soc., *Faraday Trans*. , (1994) **90** 2365 - 2374.

39. Chowdhary P, Panja S, Chatterjee A, Bhattacharya P, Chakravorti S, pH-Dependent excited-state proton transfer characteristics in 2-acetyl benzimidazole and 2-benzoyl benzimidazole in aqueous and non-aqueous media, *J. Photochemistry and Photobiology A: Chemistry,* (2005) **173** 106-113.

40. Varandas A J C, Formosinho S J, A general inter-relationship between transition-state bond extensions and the energy barrier to reaction *J. Chem. Soc., Faraday Trans. II*, (1986) **82**, 953-962.

41. Lendvay G, Bond orders from ab initio calculations and a test of the principle of bond order conservation *J. Phys. Chem.,* (1989) **93** 4422-4429.

42. Wiberg K B, Application of the pople-santry-segal CNDO method to the cyclopropylcarbinyl and cyclobutyl cation and to bicyclobutane, *Tetrahedron*, (1968) **24** 1083-1096.

43. Reed A E, Curtiss L A, Weinhold F, Intermolecular interactions from a natural bond orbital, donor-acceptor viewpoint *Chem. Rev*., (1988) **88** 899–926.

**List of Schemes:**

1. **Scheme 1:** Different structures of PCL

2. **Scheme 2:** Excited state proton transfer reaction dynamics of PCL in different environment

3. **Scheme 3:** Excited state intramolecular proton transfer ($ESI_{ra}PT$) reaction scheme of PCL in gaseous medium. Normal form(E), Transition state (TS), Zwitterionic form (Z).

4. **Scheme 4:** Excited state intermolecular proton transfer ($ESI_{er}PT$) reaction dynamics of PCL in hydroxylic medium (water).

**List of figures:**

**List of tables:**

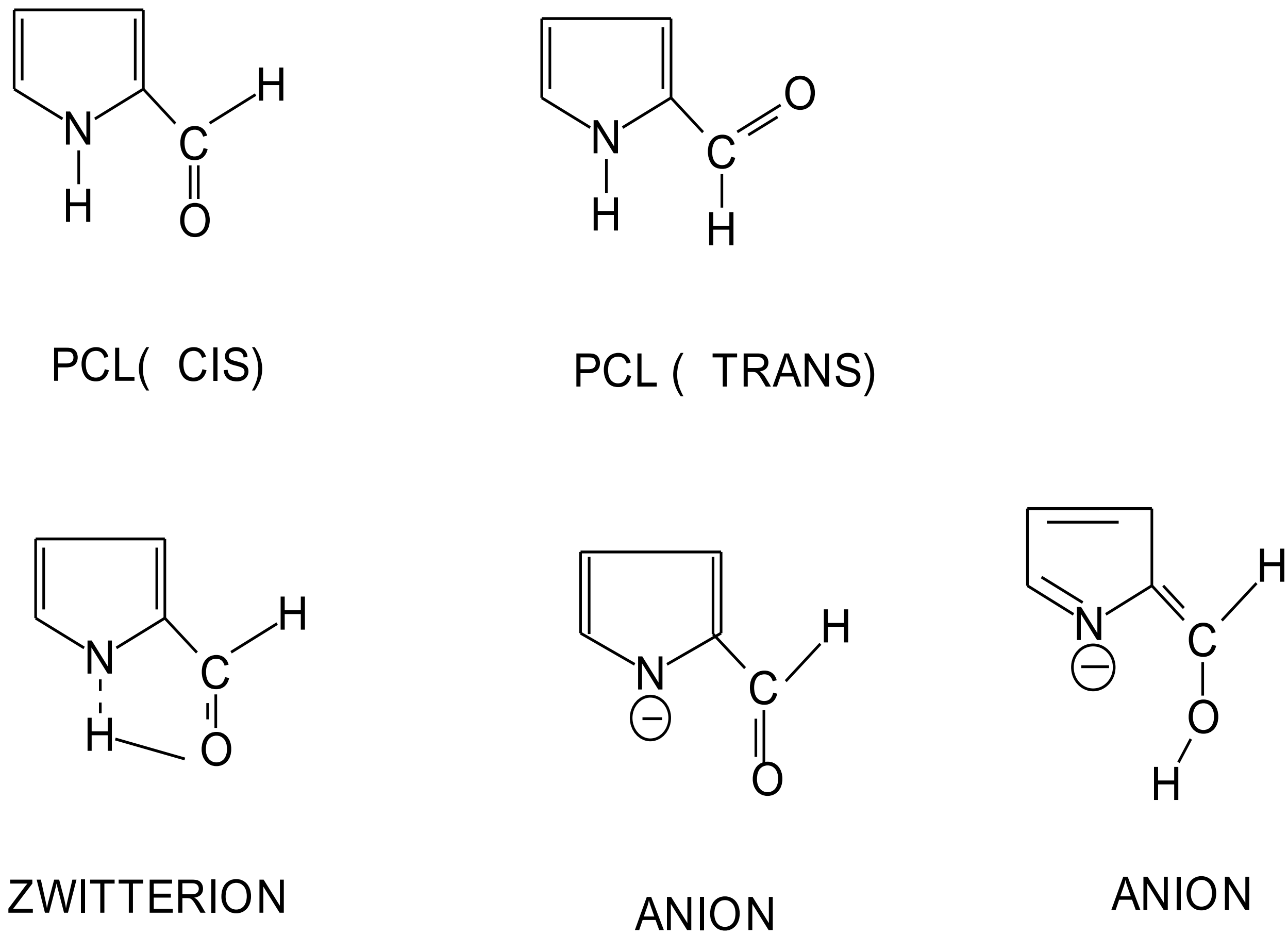

**Scheme 1:**

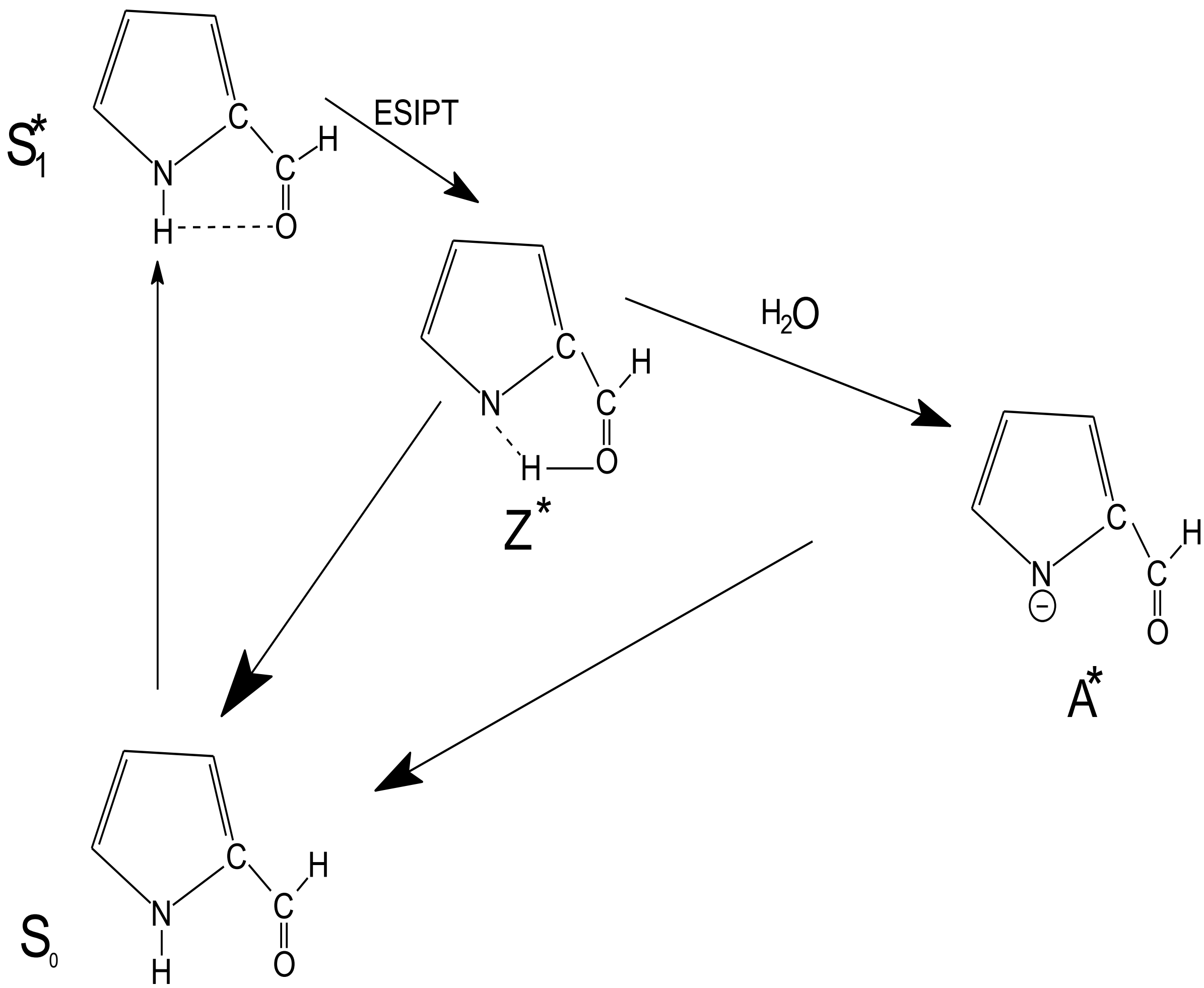

**Scheme 2:**

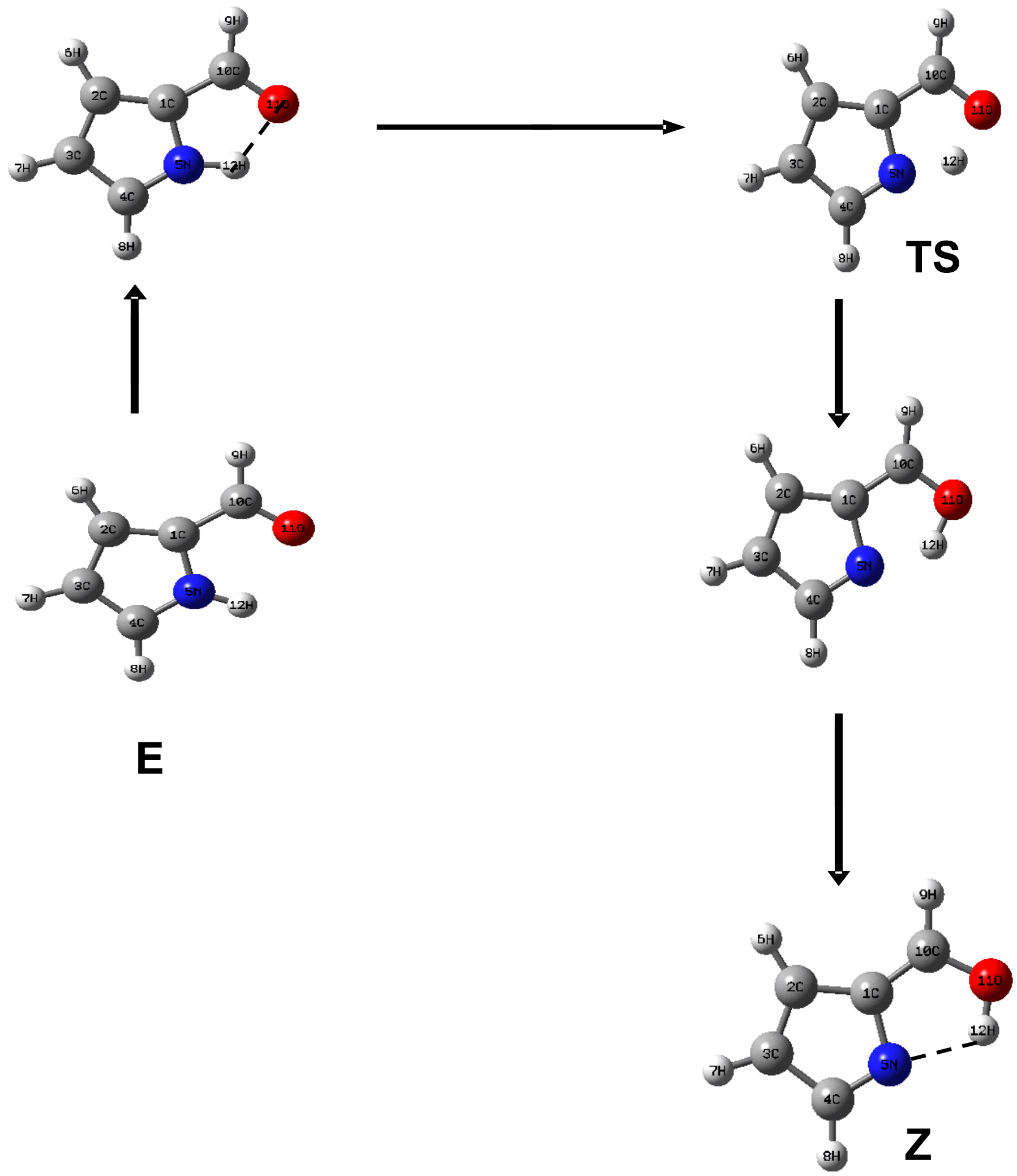

**Scheme 3:**

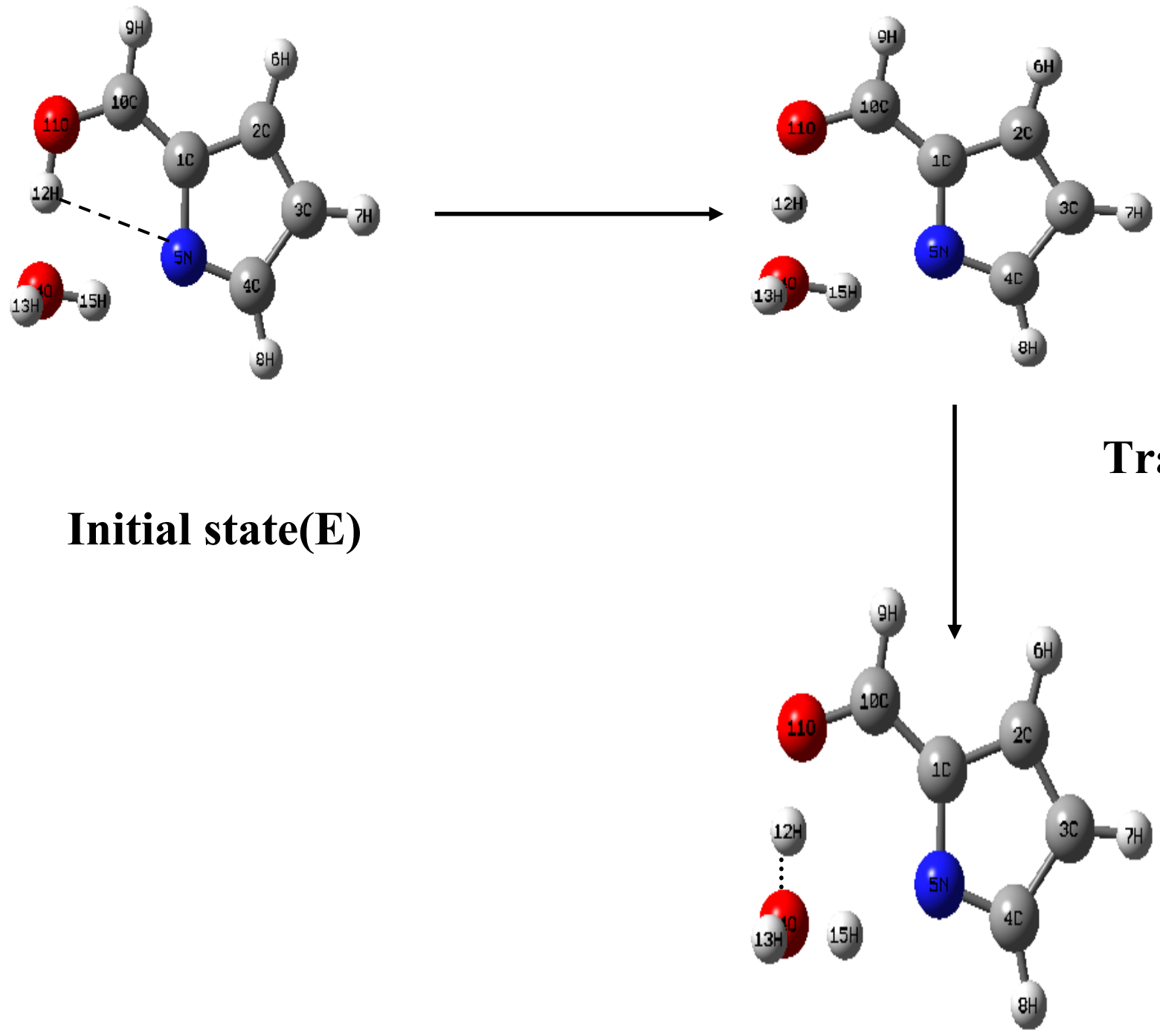

Initial state(E)

Transition state (TS)

Proton transferred state (A)

**Scheme 4:**

**Table-1:**

| **Conformer** | **Energy(Kcal/mole) in ground state** | **Dipole moment (Debye)** | **Mulliken charge density on $N_5$ (Coloumb /$m^2$)** | **Mulliken charge density on $H_6$ (Coloumb /$m^2$)** | **Enthalpy (Kcal/mol)** | **Gibbs free energy (Kcal/mol)** |
|---|---|---|---|---|---|---|
| cis (E) | -201810.98 | 3.10 | -0.64 | +0.34 | 62.91 | 40.39 |
| trans (K) | -201728.14 | 4.89 | -0.75 | +0.34 | 62.28 | 39.66 |

**Table-2:**

| Bond length, Bond Angle | PCL (cis) in ground state (DFT) | PCL (cis) in ground state ($MP_2$) | PCL (cis) in excited state | PCL (trans) in ground state (DFT) |
|---|---|---|---|---|
| B1 (C1-C2) | 1.39 | 1.39 | 1.40 | 1.39 |
| B2 (C1-N5) | 1.39 | 1.39 | 1.39 | 1.37 |
| B3 (C1-C10) | 1.43 | 1.45 | 1.40 | 1.44 |
| B4 (C2-C3) | 1.41 | 1.43 | 1.48 | 1.41 |
| B5 (C2-H6) | 1.07 | 1.08 | 1.06 | 1.07 |
| B6 (C3-C4) | 1.39 | 1.39 | 1.36 | 1.39 |
| B7 (C3-H7) | 1.07 | 1.08 | 1.06 | 1.07 |
| B8 (C4-N5) | 1.36 | 1.38 | 1.39 | 1.37 |
| B9 (C4-H8) | 1.07 | 1.08 | 1.06 | 1.07 |
| B10 (N5-H12) | 1.00 | 1.00 | 1.00 | 1.00 |
| B11 (C10-O11) | 1.25 | 1.25 | 1.26 | 1.24 |
| B12 (C10-H9) | 1.10 | 1.10 | 1.07 | 1.10 |
| B13 (H12-O11) | 2.60 | 2.63 | 2.31 | 2.52 (H9-H12) |
| A1 (C2-C1-N5) | 107.1 | 107.3 | 104.7 | 106.8 |
| A2 (C2-C1-C10) | 132.1 | 131.6 | 138.3 | 131.1 |
| A3 (N5-C1-C10) | 120.7 | 121.0 | 116.9 | 122 |
| A4 (C1-C2-C3) | 107.5 | 107.5 | 107.2 | 107.9 |
| A5 (C1-C2-H6) | 125.4 | 125.8 | 127.1 | 124.4 |
| A6 (C3-C2-H6) | 126.9 | 126.6 | 125.5 | 126.7 |
| A7 (C2-C3-C4) | 107.4 | 107.4 | 108.3 | 107.5 |
| A8 (C2-C3-H7) | 126.8 | 126.6 | 124.8 | 126.7 |
| A9 (C4-C3-H7) | 125.7 | 125.9 | 126.8 | 125.6 |
| A14 (C1-N5-H12) | 122.3 | 123.1 | 118.8 | 124.9 |
| A15 (C4-N5-H12) | 127.8 | 126.9 | 127.4 | 125.3 |
| A16 (C1-C10-O11) | 123.3 | 122.9 | 119.8 | 124.9 |
| A17 (C1-C10-H9) | 115.8 | 116.5 | 118.5 | 115.3 |
| A18 (O11-C10-H9) | 120.7 | 120.4 | 121.6 | 119.7 |

**Table-3:**

| Step no. | Energy (Hartree) | Reaction coordinate (A°) | Bond length (A°) | Internal force constant (RMS) |
|---|---|---|---|---|
| 1 | -321.52 | -1.99 | 1.90 | 0.00 |
| 4 | -321.52 | -1.69 | 1.84 | 0.09 |
| 8 | -321.51 | -1.29 | 1.77 | 0.03 |
| 12 | -321.51 | -0.89 | 1.67 | 0.09 |
| 16 | -321.50 | -0.49 | 1.53 | 0.07 |
| 20 | -321.49 | -0.09 | 1.34 | 0.06 |
| 24 | -321.50 | 0.29 | 1.16 | 0.05 |
| 28 | -321.52 | 0.69 | 1.03 | 0.05 |
| 32 | -321.53 | 1.09 | 1.00 | 0.08 |
| 36 | -321.54 | 1.49 | 0.99 | 0.09 |
| 40 | -321.54 | 1.89 | 0.99 | 0.08 |

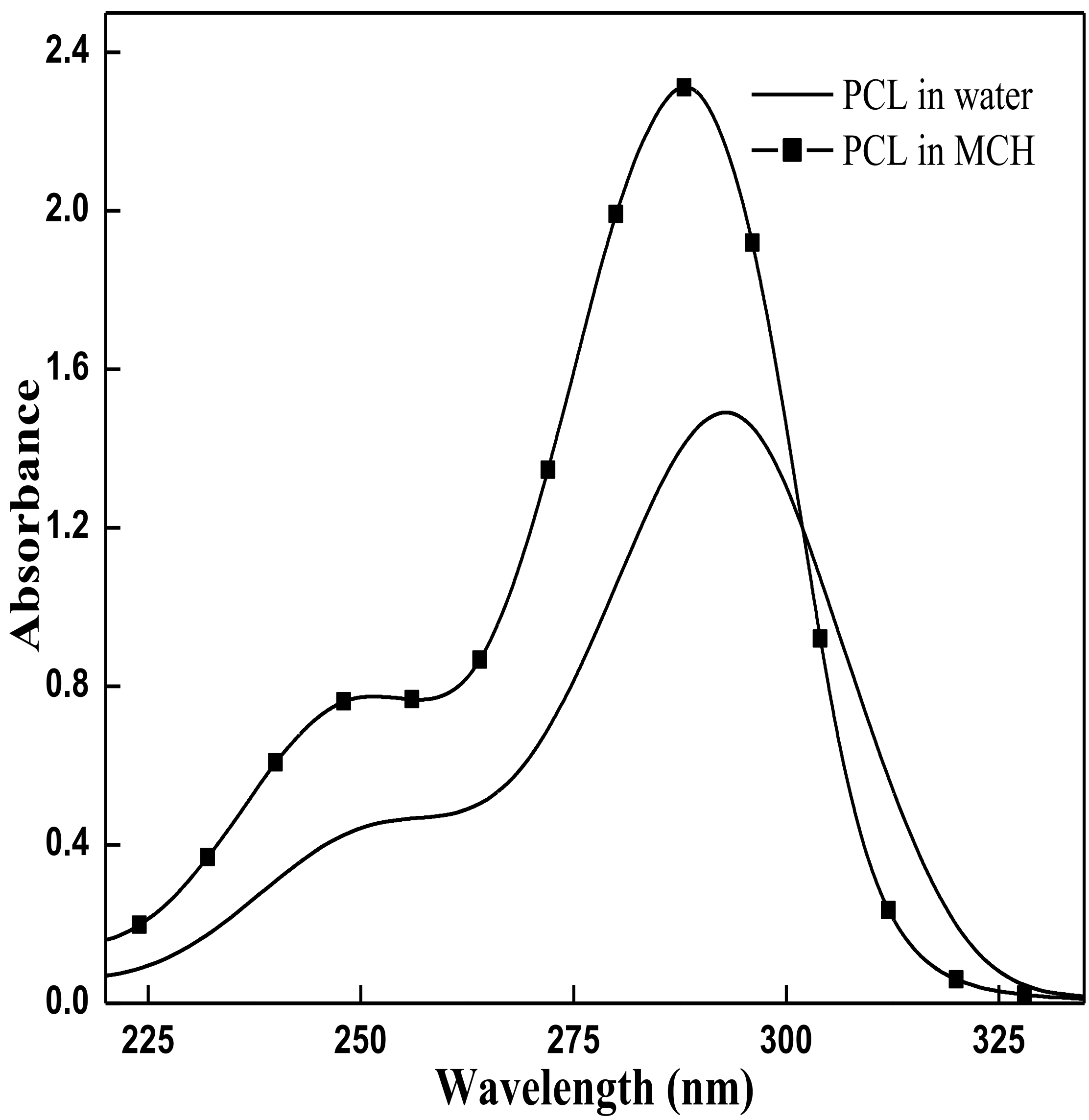

PCL in water
PCL in MCH
Absorbance
Wavelength (nm)
0.0
0.4
0.8
1.2
1.6
2.0
2.4
225
250
275
300
325

**Figure-1:**.

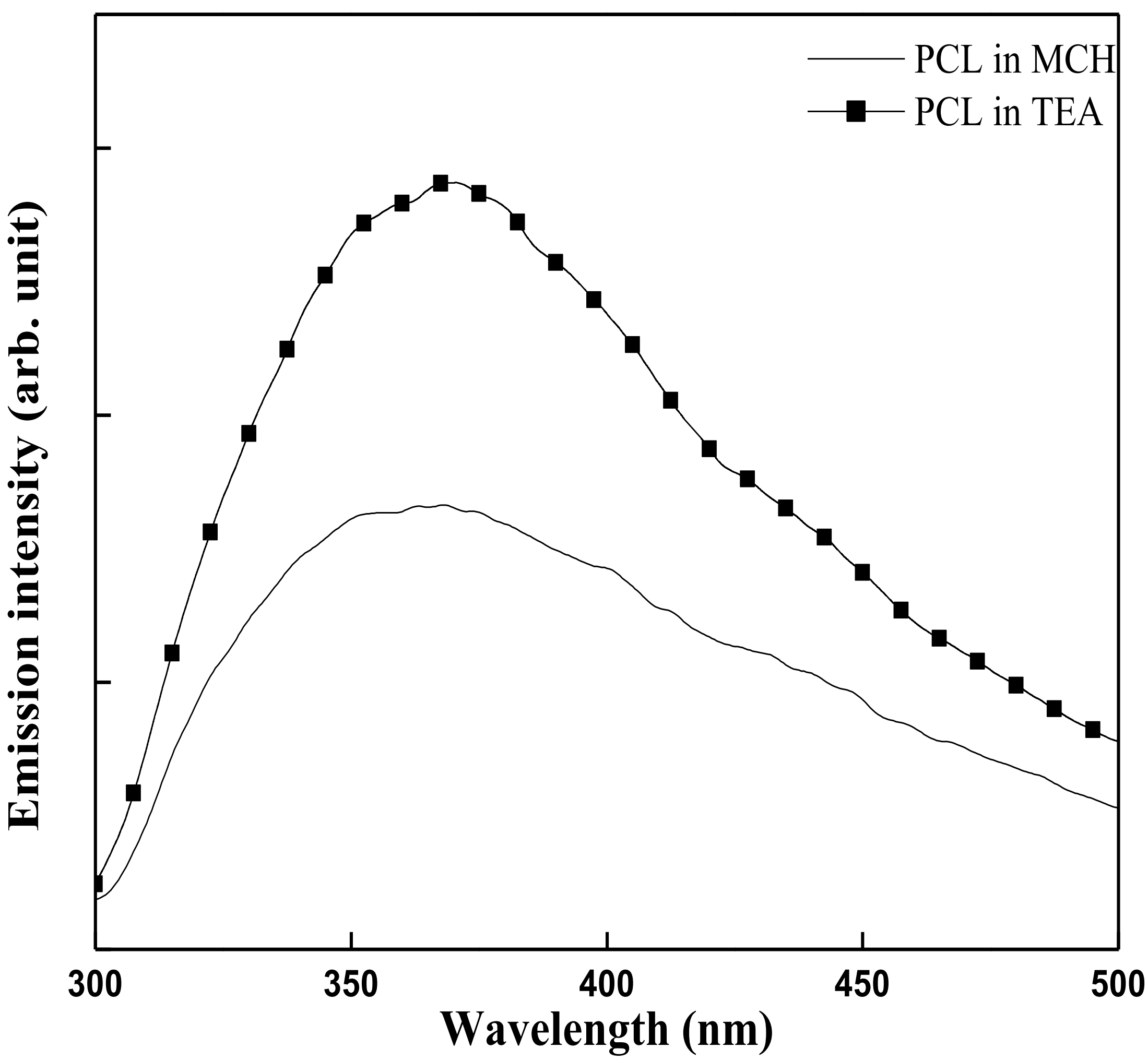

PCL in MCH
PCL in TEA
Emission intensity (arb. unit)
300
350
400
450
500
Wavelength (nm)

**Figure-2:**

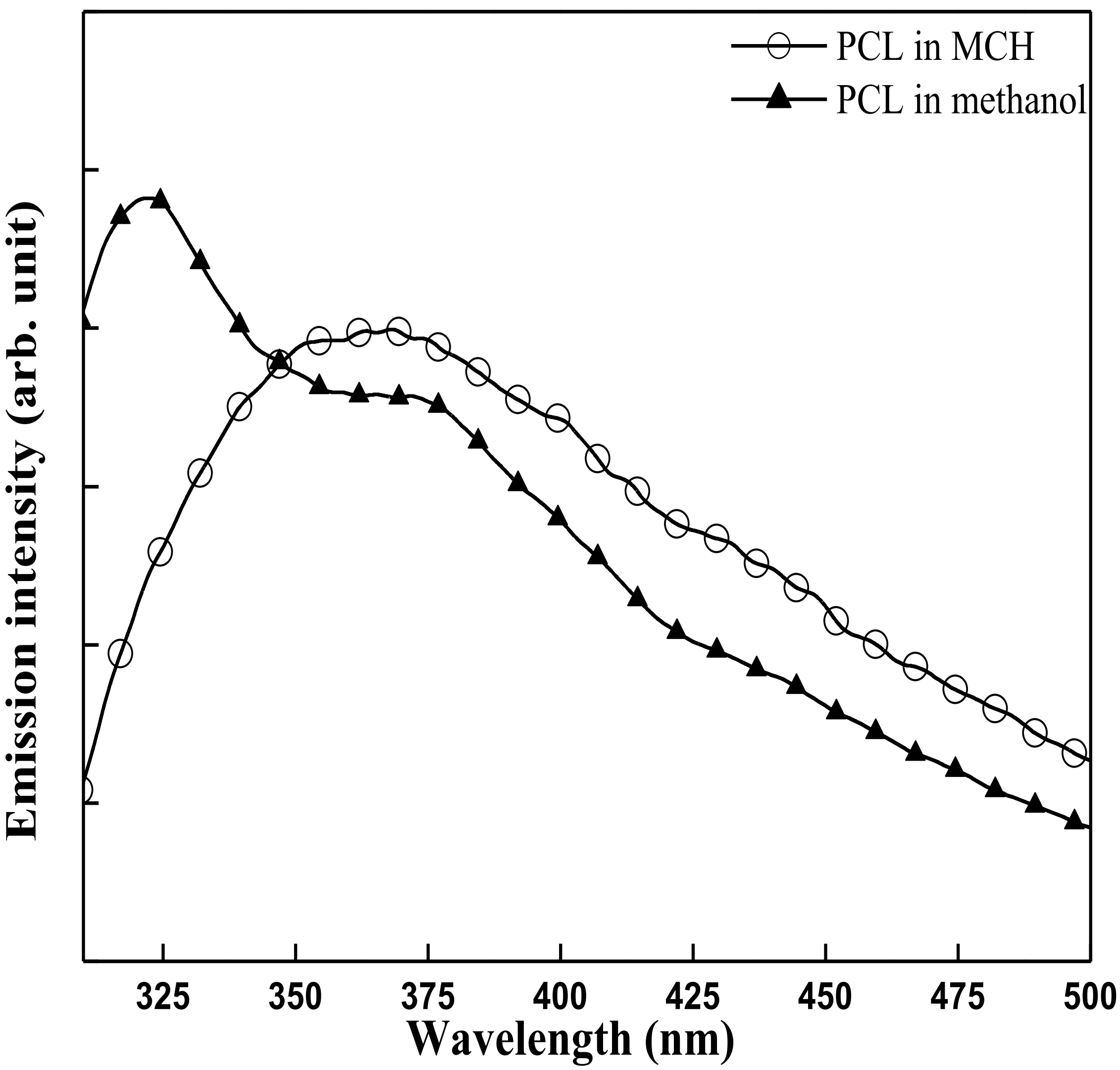

PCL in MCH
PCL in methanol
Emission intensity (arb. unit)
325
350
375
400
425
450
475
500
Wavelength (nm)

**Figure-3:**

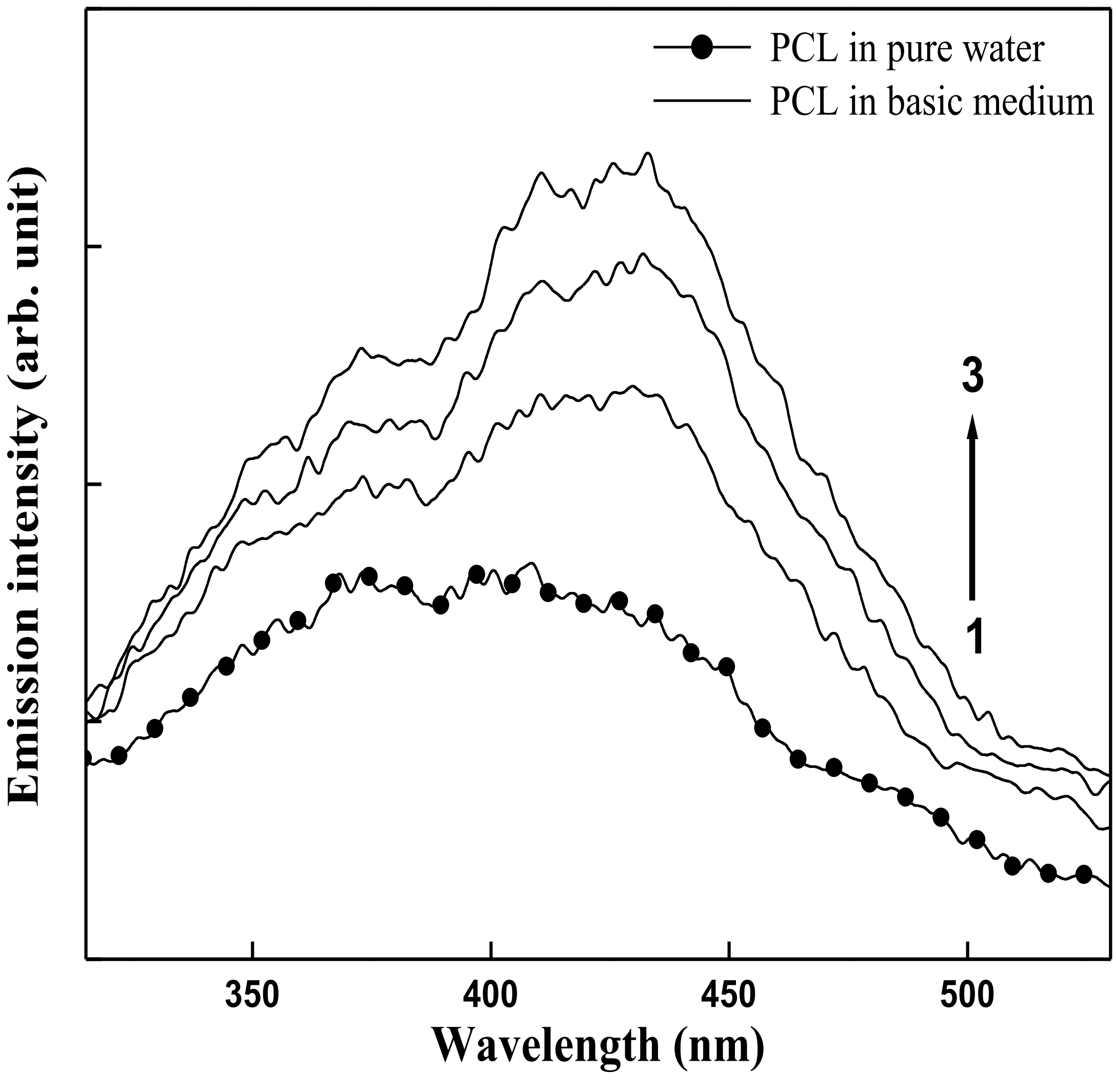

PCL in pure water
PCL in basic medium
3
1
Emission intensity (arb. unit)
350
400
450
500
Wavelength (nm)

**Figure-4:**

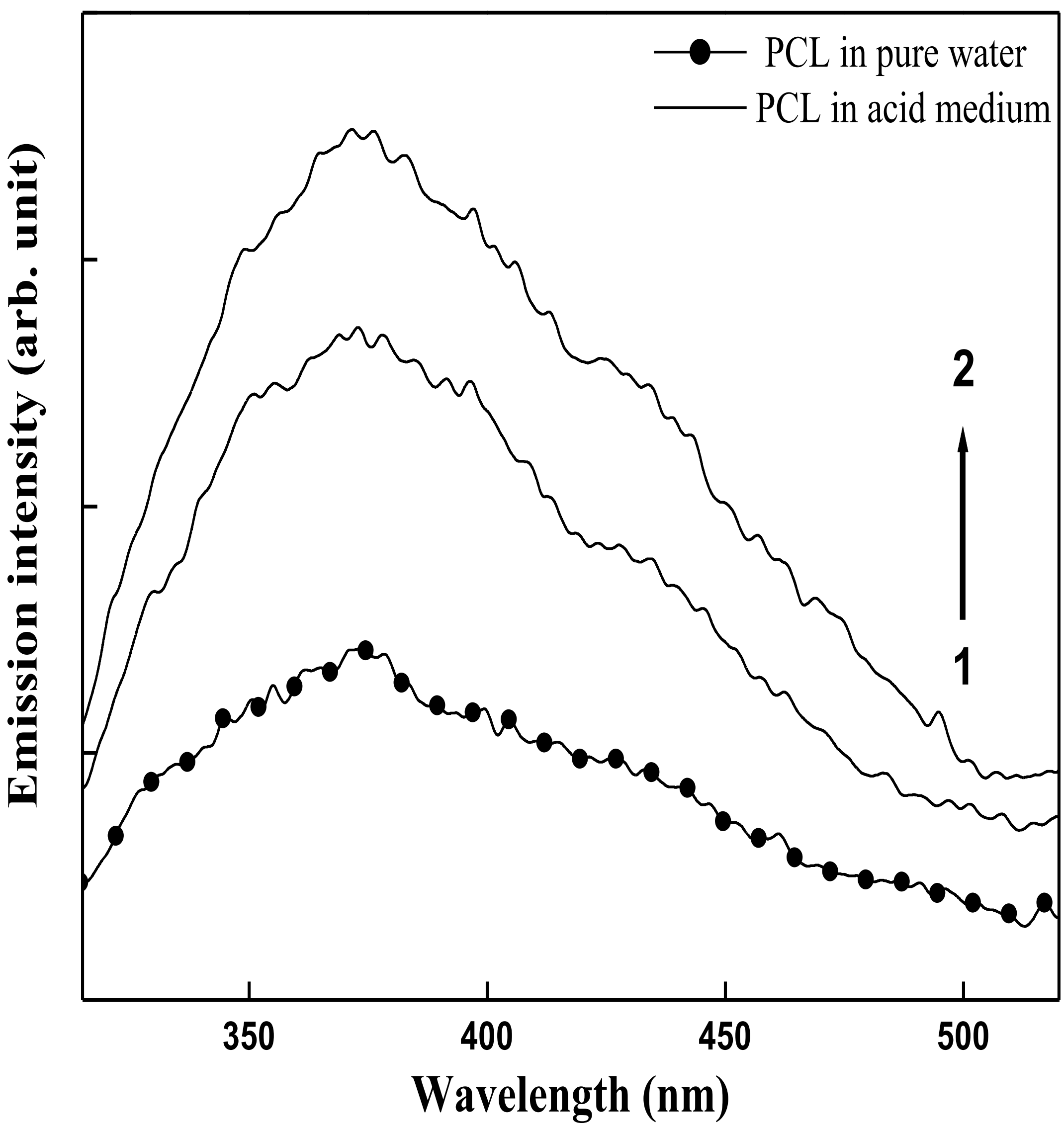
PCL in pure water
PCL in acid medium
2
1
Emission intensity (arb. unit)
350
400
450
500
Wavelength (nm)

**Figure-5:**

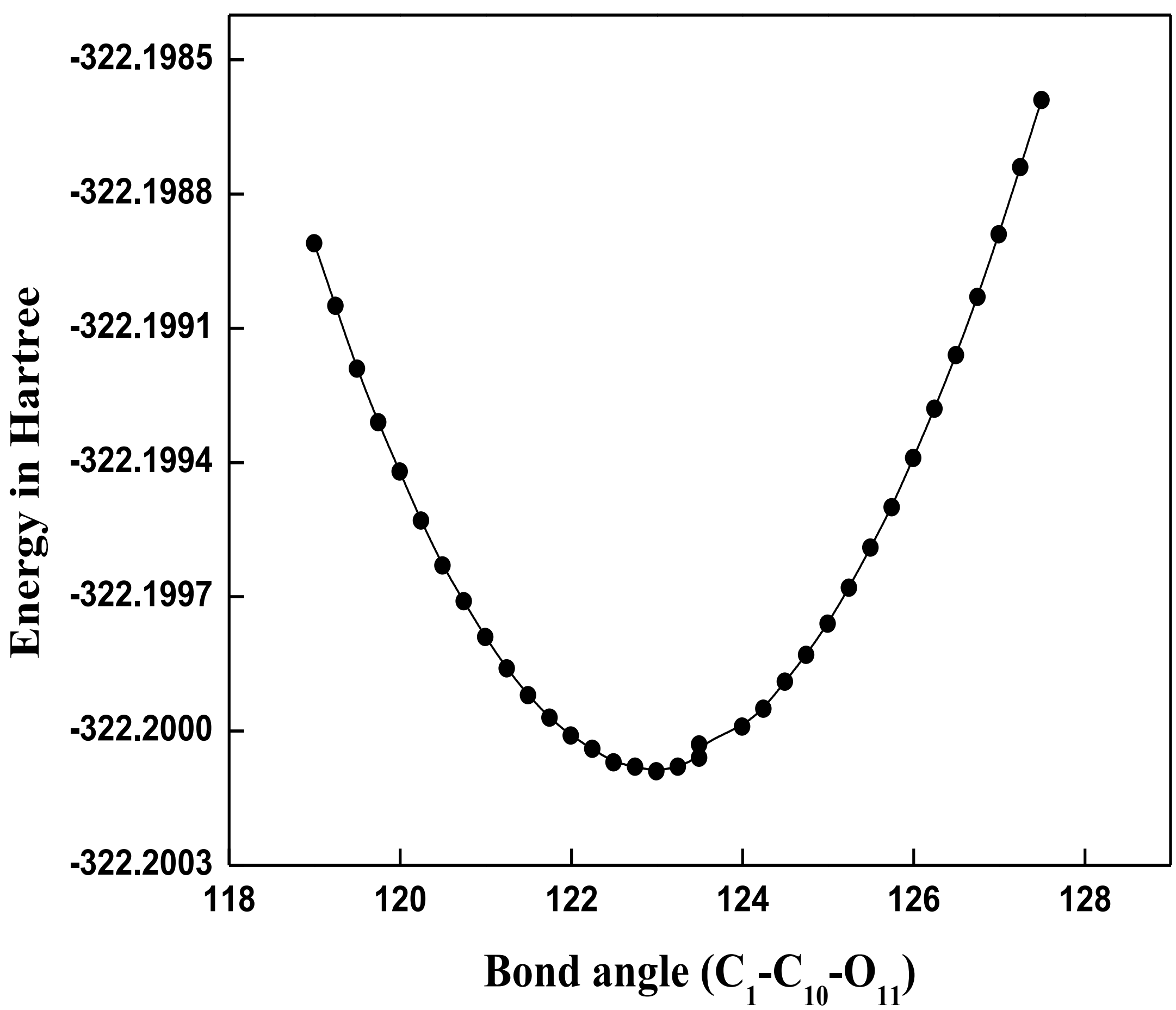
Energy in Hartree
-322.1985
-322.1988
-322.1991
-322.1994
-322.1997
-322.2000
-322.2003
118
120
122
124
126
128
Bond angle ($C_1$-$C_{10}$-$O_{11}$)

**Figure-6:**

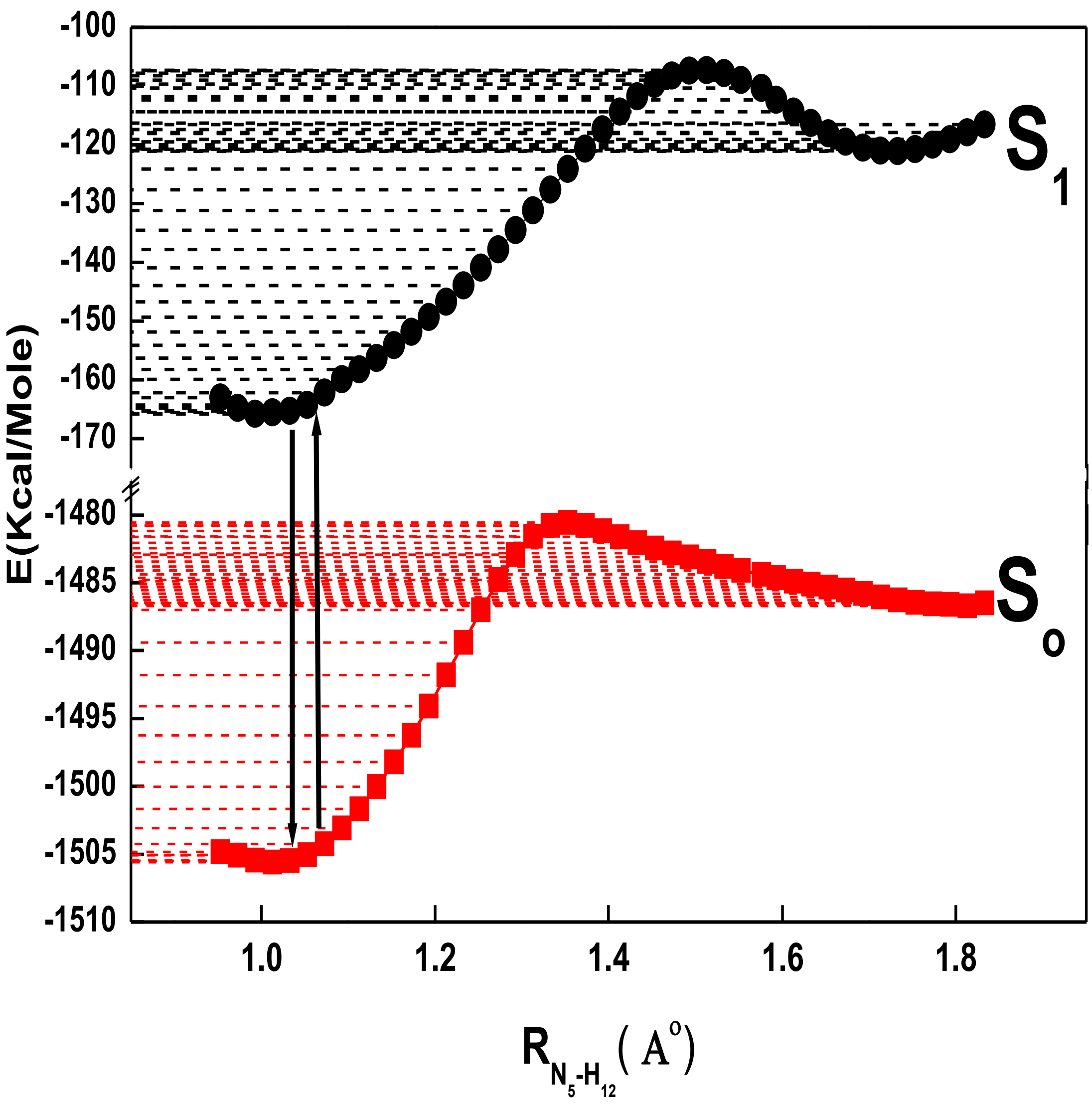

-100
-110
-120
-130
-140
-150
-160
-170
-1480
-1485
-1490
-1495
-1500
-1505
-1510
E(Kcal/Mole)
S1
So
1.0
1.2
1.4
1.6
1.8
R N5-H12 ( A° )

**Figure-7:**

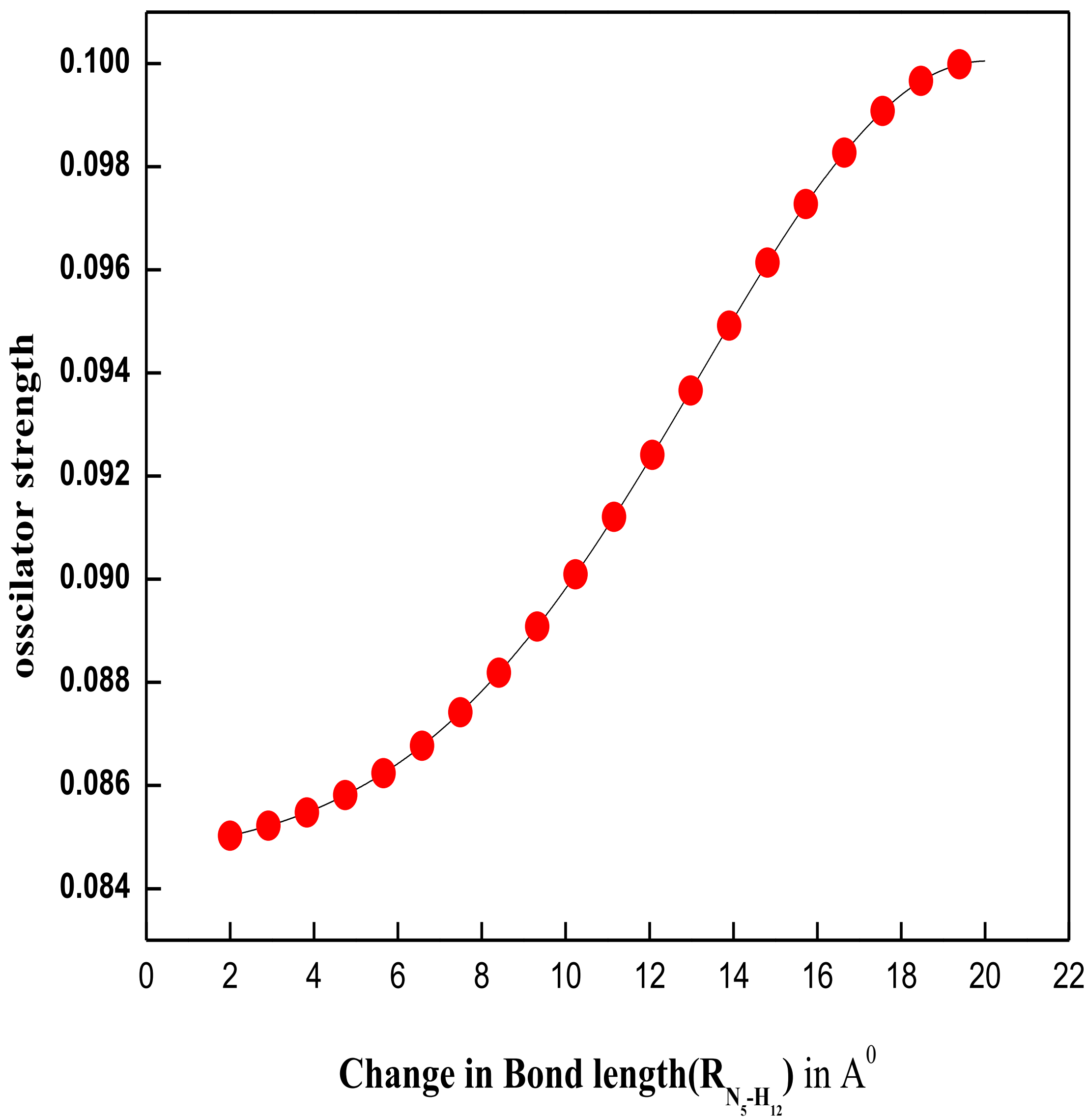
osscilator strength
0.100
0.098
0.096
0.094
0.092
0.090
0.088
0.086
0.084
0
2
4
6
8
10
12
14
16
18
20
22
Change in Bond length($R_{N_5\text{-}H_{12}}$) in $A^0$

**Figure-8:**

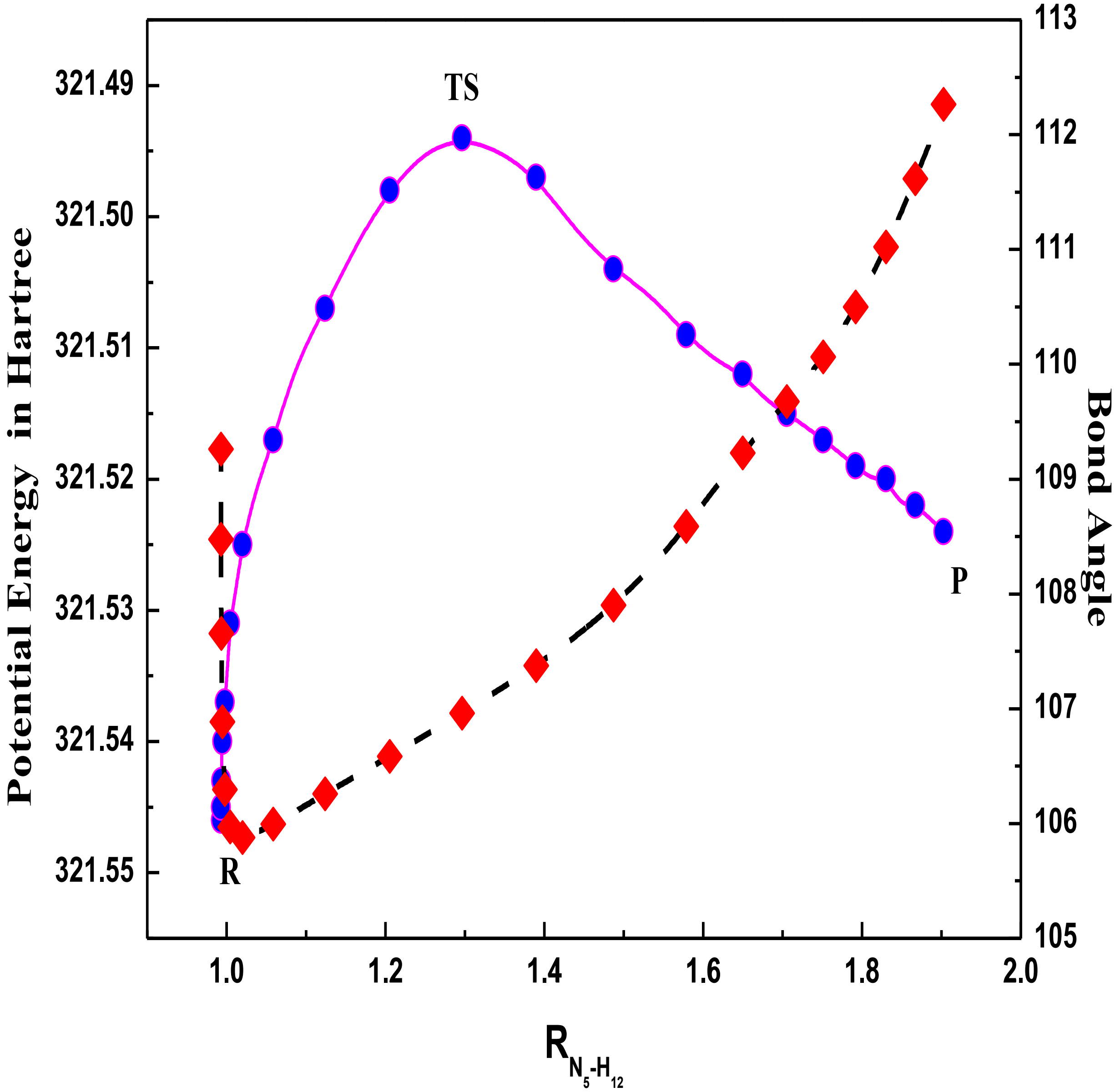

TS
R
P
321.49
321.50
321.51
321.52
321.53
321.54
321.55
113
112
111
110
109
108
107
106
105
1.0
1.2
1.4
1.6
1.8
2.0
Potential Energy in Hartree
Bond Angle
R N5-H12

**Figure-9:**

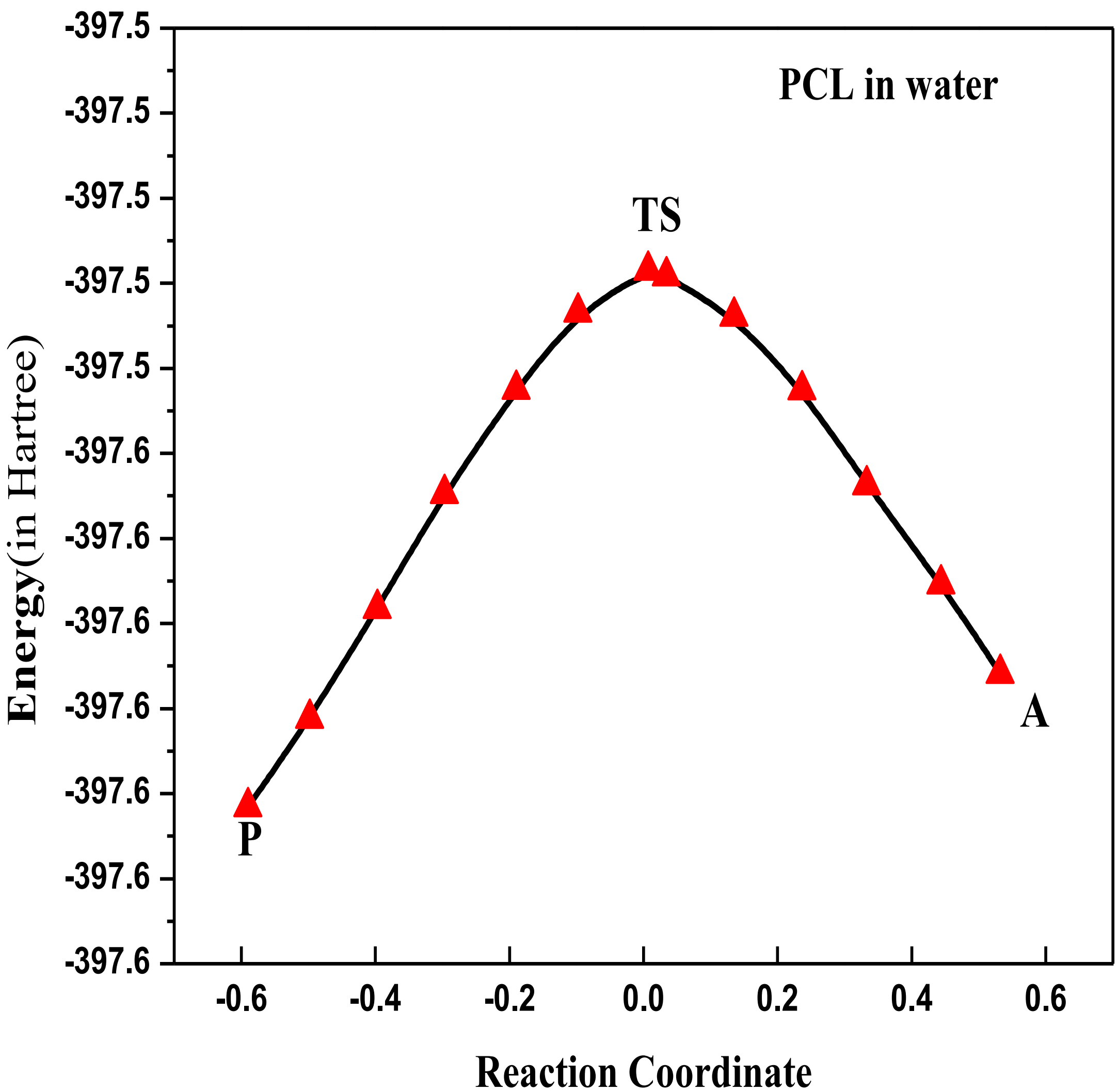
-397.5
-397.5
-397.5
-397.5
-397.5
-397.6
-397.6
-397.6
-397.6
-397.6
-397.6
-397.6
PCL in water
TS
P
A
Energy(in Hartree)
-0.6
-0.4
-0.2
0.0
0.2
0.4
0.6
Reaction Coordinate

**Figure-10:**